# Real-time Observation of Thermal Surface Recovery in SrVO$_3$

Amit Cohen[1], Jonathan Ludwick[2,3], Ward Yahya[1], Maria Baskin[1], Lishai Shoham[1], Tyson C. Back[2], Lior Kornblum[*1]

[1]Andrew and Erna Viterbi Department of Electrical and Computer Engineering, Technion—Israel Institute of Technology, Haifa 32000-03, Israel

[2]Air Force Research Laboratory, WPAFB, 2179 12th Street, B652/R122, Dayton, OH 45433-7718, USA

[3]UES BlueHalo Company, 4401 Dayton-Xenia Rd, Dayton, OH 45432, USA

*Corresponding Author: liork@technion.ac.il

## ABSTRACT

SrVO$_3$ (SVO), a model correlated metal and a promising transparent conducting oxide, develops a several-nanometer-thick near-surface region (NSR), rich in V$^{5+}$ species under ambient conditions. This oxidized layer obscures the intrinsic correlated-metallic V$^{4+}$ character and limits both fundamental studies of the physics and the material's integration into electronic devices. Here, we demonstrate a direct and controllable approach for recovering the metallic SVO surface by thermally reducing the NSR under ultra-high vacuum. Real-time *in-situ* X-ray photoelectron spectroscopy (XPS) reveals a sharp transformation from a V$^{5+}$-dominated surface to mixed valence states, dominated by V$^{4+}$, and a recovery of its metallic character. *Ex-situ* X-ray diffraction (XRD), atomic force microscopy (AFM), and high-resolution scanning electron microscopy (HR-SEM) suggest that this transformation is accompanied by mass redistribution and partial oxygen loss, leading to nanoscale surface reorganization and modest lattice expansion. While thermodynamic considerations motivate evaluation of a V$_2$O$_5$ volatilization pathway, the combined experimental evidence instead points toward a predominantly structural surface reorganization. These findings establish a practical method for obtaining predominantly V$^{4+}$ SVO surfaces without protective capping layers, a capability that expands the utility of SVO for advanced spectroscopies, interface engineering, and oxide-electronics device integration.

## I. INTRODUCTION

Transition metal oxides (TMOs), or functional oxides, host an abundance of interesting physical phenomena, particularly those arising from electron correlation. These materials therefore are attracting significant attention for both their fundamental physics[1,2] and for harnessing these phenomena into emerging technological devices.[3–5] Surfaces and interfaces of TMOs are a special focal point, both for fundamental science and for technology.[6–9]

Over the past decade, the role of surface and interface chemistry has become increasingly crucial for some important TMOs. Many TMOs have more than one metastable oxidation state of the transition metal ion (e.g. the B cation in the $ABO_3$ perovskite archetype). While this feature hosts many elaborate and interesting electronic and magnetic behaviors, it also makes some TMOs unstable in air, where the 'flexibility' in valence can manifest in paths for more stable oxidation states, which are different from the bulk. Such surface chemical changes can be detrimental for functional devices and therefore call for careful consideration when analyzing physical properties.[10]

A useful case study for electron correlation, as well as surface stability, is strontium vanadate, $SrVO_3$ (SVO). It is a cubic, strongly correlated $d^1$ metal,[11–15] and a case study of filling-controlled[16–18] and bandwidth-controlled[17,19] Mott physics. It receives significant attention as a potential transparent conducting oxide[20–22] (TCO), a crucial component of optoelectronic devices, and specifically of photovoltaics. In this context, SVO offers a significant advantage, being made from earth-abundant elements, in contrast to the current TCO champion of indium tin oxide (ITO), therefore offering potential advantages in photovoltaic scalability and cost-effectiveness.

In a simple ionic picture, V is expected to hold the +4 oxidation state in bulk SVO. V can also host the +3 oxidation state (e.g. in $LaVO_3$), and the +5 (e.g. in $V_2O_5$). It was found that air-exposed SVO surfaces readily oxidize to a predominately $V^{+5}$ state,[23–25] which can extend to some ~10 unit cells (~4 nm). The resulting surface can be complex chemically and structurally, and thorough characterization has revealed $Sr_3V_2O_8$ as a common surface phase;[26–28] furthermore, it was shown that this phase can be selectively etched in water and water vapor.[26,29] Alternatively, it was shown that a few-nm thick capping layers of typically-amorphous oxides, are highly effective in preventing the formation of $V^{+5}$ surface phases,[23] while preserving a bulk-like $V^{+4}$ SVO under the cap.[30]

In this work we describe a new approach for restoring air-exposed SVO surfaces to their predominately $V^{+4}$ state. This reduction is achieved by annealing in ultra-high vacuum (UHV). We continuously monitor the chemical state while ramping up the temperature using *in-situ* X-ray photoelectron spectroscopy (XPS) and observe the onset and evolution of the reduction from a $V^{+5}$ dominated surface to its desired $V^{+4}$, restoring stochiometric SVO at the surface.

This approach enables potential advantages for surface-sensitive characterization techniques, such as angle-resolved photoelectron spectroscopy (ARPES), by recovering a well-defined correlated-metallic surface following ambient exposure. Moreover, this approach opens pathways for *ex-situ* deposition of additional materials and broadens the engineering landscape of these materials into electronic and optoelectronic devices.

## II. EXPERIMENTAL

Film growth: Epitaxial SVO films were grown on (001) $(LaAlO_3)_{0.3}(Sr_2TaAlO_6)_{0.7}$ (LSAT) substrates using oxide molecular-beam epitaxy (MBE). Growth followed the co-deposition procedure described elsewhere.[31]

Real-time *in-situ* XPS: measurements were performed continuously under UHV with a base pressure of ~$1 \times 10^{-10}$ Torr. High-resolution core-level spectra were acquired with a pass energy of 20 eV. For the temperature-dependent XPS experiment, the sample was heated from 670 K to 1052 K and cooled back to 812 K at a rate of ~2 K/min (Fig. S1) while continuously acquiring V 2p and O 1s spectra (Fig. S2). The temperature was monitored using an optical pyrometer (emissivity 0.9).

*Ex-situ* characterization: Following the *in-situ* XPS measurements, the sample was removed from vacuum and characterized *ex-situ*. X-ray diffraction (XRD) was performed in a 2θ/ω geometry using a Rigaku SmartLab. Surface morphology was imaged using an Asylum Research/Oxford Instruments Cypher ES atomic force AFM operated in tapping mode, as well as a Zeiss Ultra-Plus high-resolution scanning electron microscopy (HR-SEM) equipped with an in-column, energy-selective backscattered-electron detector (EsB) and a conventional secondary-electron (SE) detector. *Ex-situ* XPS measurements were acquired with a Thermo VG Scientific Sigma Probe system. A pass energy of 20 eV and surface normal collection angle were employed, and a monochromated Al Ka source (1468.6 eV) was used. Binding energies were calibrated by *in-situ* measurements.

## III. RESULTS AND DISCUSSION

## A. Spectroscopic Evolution of Thermal Surface Recovery

XPS spectra of the SVO film acquired at the beginning ("before") and at the end ("after") of the thermal treatment reveal a pronounced transformation in the chemical and electronic states of vanadium and strontium (Fig. 1). Prior to heating, the V $2p_{3/2}$ peak (Fig. 1a) is dominated by a $V^{5+}$ component (~518 eV) originating from an oxidized near-surface region (NSR), with only a minor contribution from the buried $V^{4+}$ (~517 eV) stoichiometric SVO (Fig. S3), in agreement with previous work.[23] Following the treatment, the V $2p_{3/2}$ line shape becomes markedly more complex, whereas *ex-situ* measurements after air exposure show a recovery of the pre-treatment line shape. This behavior of the post-treatment *in-situ* spectrum is consistent with the known electronic structure of SVO, where the complex $V^{4+}$ spectral signature arises from non-local charge-transfer, characteristic of a correlated metallic state.[30,32] The spectrum further reveals the coexistence of additional oxidation states, indicating partial reduction of the surface region.

Previously acquired spectra of $V^{3+}$ ($LaVO_3$), $V^{4+}$ (stoichiometric SVO), and $V^{5+}$ (oxidized SVO NSR)[23,30] were superimposed on the post-treatment data (Fig. 1b). It is important to note that ascribing formal oxidation states in these scenarios is an oversimplification,[33] which we do for clarity. Crucially, the $V^{4+}$ contribution is defined by the characteristic V 2p line shape of stoichiometric SVO, arising from non-local charge-transfer processes and strong electronic correlations.[30,34] The stoichiometric SVO $V^{4+}$ state comprises roughly 65% of the spectrum, where the $V^{3+}$ and $V^{5+}$ spectral weight is smaller. This quantitative analysis suggests that the newly formed surface is predominantly stoichiometric SVO, while minor fractions of oxidized ($V^{5+}$) and reduced ($V^{3+}$) species correspond respectively to remnants of the initial NSR and to reduction processes occurring during the treatment.

The valence band spectrum evolves from a featureless gap near the Fermi level to a distinct metallic peak (Fig. 1c), consistent with the re-emergence of the metallic SVO phase at the surface. The Sr 3d peak likewise starts with a broad feature dominated by a surface phase at a higher binding energy compared to SVO.[23] Following the treatment, the Sr spectrum is dominated by the bulk SVO doublet,[23] suggesting that the SVO phase has re-emerged at the surface. Altogether, these spectral changes indicate the removal of a significant portion of the NSR, and the appearance of SVO near the surface.

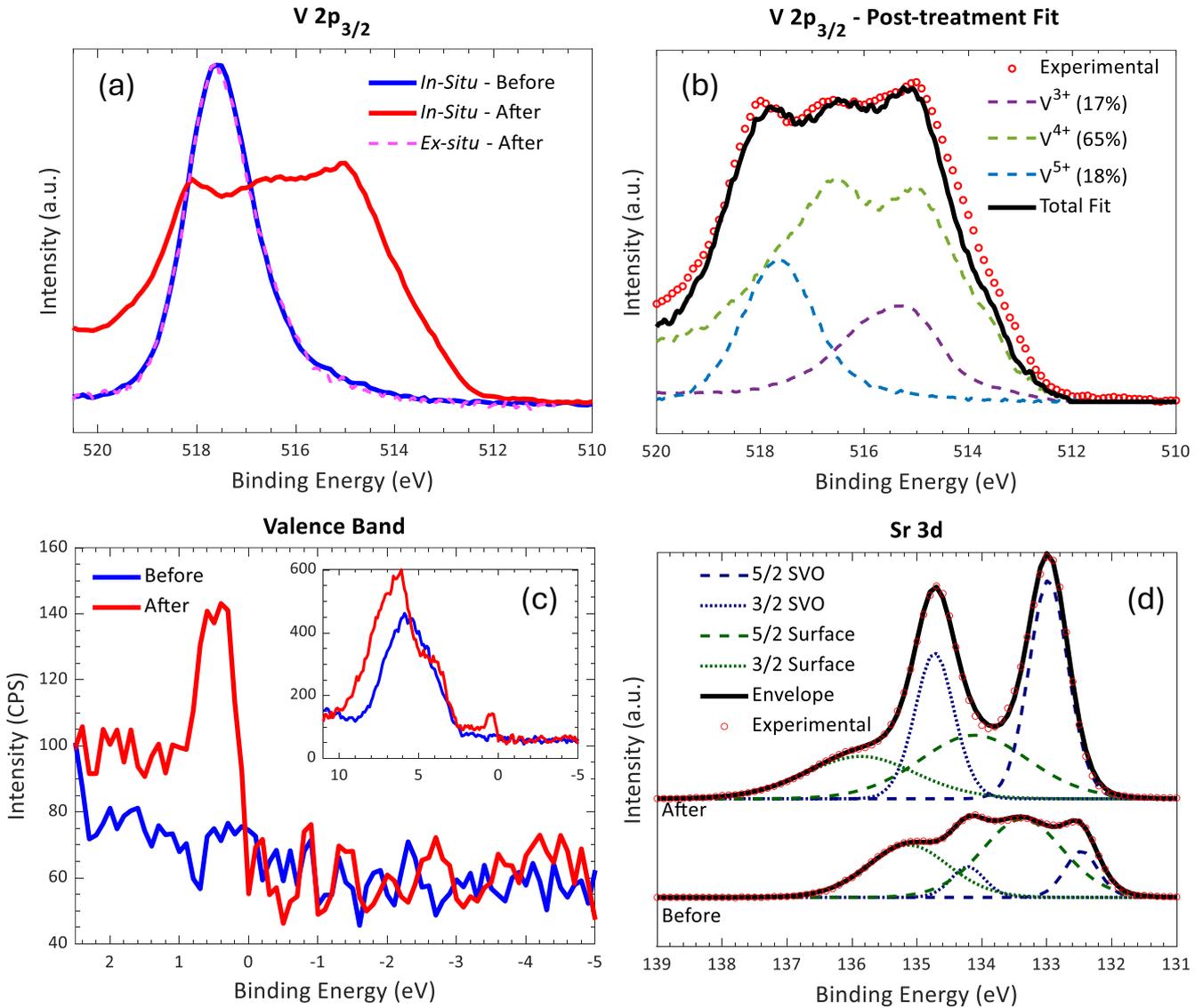

**Figure 1.** *In-situ* XPS spectra of the SVO film, before and after thermal treatment (intensities shown as received). (a) V $2p_{3/2}$, with an additional *ex-situ* spectrum after air exposure, which is scaled for visual comparison (full range in Fig. S4). (b) Fitting of the post-treatment V $2p_{3/2}$ spectrum using reference components of $V^{3+}$ (LaVO$_3$), $V^{4+}$ (stoichiometric SVO), and $V^{5+}$ (oxidized SVO NSR).[30] The fitted composition corresponds to ≈ 65% $V^{4+}$, 17% $V^{3+}$, and 18% $V^{5+}$. (c) Valence band spectra, inset shows a wider range corresponding to the boxed region in the main panel. (d) Sr 3d.

The evolution of the surface chemistry during the thermal treatment was continuously monitored *in-situ*. To emphasize the transformation, Fig. 2 presents only the V 2p region and the temperature range around the transition (for the full spectral and temperature ranges, see Fig. S2 and discussion therein). Near 970 K, the V 2p exhibits a well-defined onset of line-shape evolution from a $V^{5+}$ dominated feature at 518 eV into the more complex line shape associated with mixed $V^{3+}$, $V^{4+}$, and $V^{5+}$ states, accompanied by a pronounced spectral weight shift towards lower binding energies (Fig. 2b), signaling the recovery of the correlated-metallic surface electronic structure.

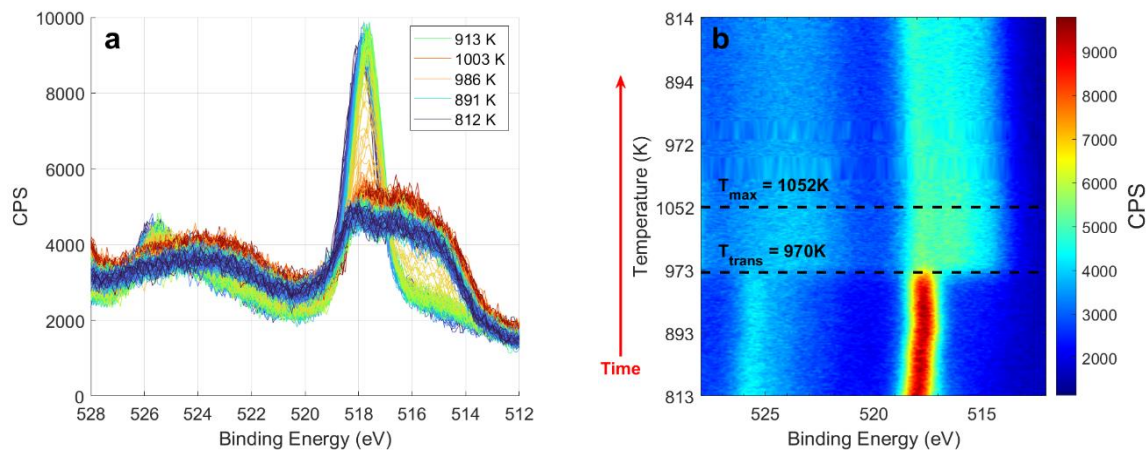

**Figure 2.** (a) XPS spectra of V 2p taken in continuous heating from 813 K to 1052K, and down back to 815 K. (b) Heat map of the XPS spectra, y axis shows temperatures chronologically. $T_{trans}$ represents the temperature where the transition from a $V^{5+}$-dominated surface to mixed valence states occurs.

## B. *Ex-situ Characterization: Physical Mechanisms and Chemical Pathways*

While *in-situ* XPS reveals a pronounced chemical transformation of the NSR, it is important to understand how this process impacts the SVO film. *Ex-situ* structural characterization is employed to assess whether the perovskite phase, crystallographic integrity, and film continuity are preserved following thermal treatment. In addition, these measurements provide insight into the structural and morphological *physical mechanisms* governing the surface evolution accompanying the re-emergence of stoichiometric SVO. Several physical surface-evolution mechanisms may contribute to the observed evolution. These include NSR thermally-driven oxygen loss (under UHV), partial detachment, evaporation, crystallization and islanding involving surface mass transport. Such processes would expose the underlying SVO bulk film and could account for the observed spectral evolution.

The film's structural and morphological changes were examined using XRD and AFM before and after the treatment, and HR-SEM afterwards. It is important to note that these are *ex-situ* measurements, meaning that the surface is re-oxidized (Fig. 1a). The XRD patterns (Fig. 3a) confirm that the film retains its perovskite structure and high crystallinity, showing a small lattice expansion (from 3.817 Å to 3.831 Å), possibly arising from oxygen vacancies.[35] The thickness fringes become less pronounced after heating, indicating some increased disorder and surface roughening. In addition, the lower fringe frequency corresponds to a ~1.5 nm decrease in film thickness, smaller than the typical thickness of the oxidized NSR,[23] suggesting limited material redistribution rather than complete removal of this layer.

AFM images support this interpretation: the initially smooth surface, characterized by a low root-mean-square roughness (Rq = 0.16 nm; Fig. 3b), evolves into a morphology containing nanoscale islands up to 30 nm tall, while the roughness of the background between them increases to ~2 nm (Fig. 3c,d). Particle coverage analysis indicates that this background region constitutes the majority of the surface area (Fig. S5), in agreement with the *in-situ* XPS results, which show $V^{4+}$ as the dominant chemical state after the thermal treatment (Fig. 1b). On this basis, the exposed background region is attributed to stoichiometric SVO.

Consistent behavior is observed in HR-SEM, with the surface displaying a dewetting-like morphology where nanoscale islands (of comparable size to those observed by AFM) are surrounded by recessed regions, indicative of localized mass transport (Fig. 3e). The islands appear darker in the EsB-detector image (Fig. 3f), consistent with lower Z-contrast,[36] suggesting a compositionally distinct phase.

Together, these results indicate a preservation of the SVO perovskite structure and its crystallographic integrity. At the same time, the observations point to a set of thermally driven physical mechanisms, involving oxygen desorption and surface mass transport, leading to partial redistribution of material and significant surface reorganization, rather than complete detachment of the oxidized overlayer.

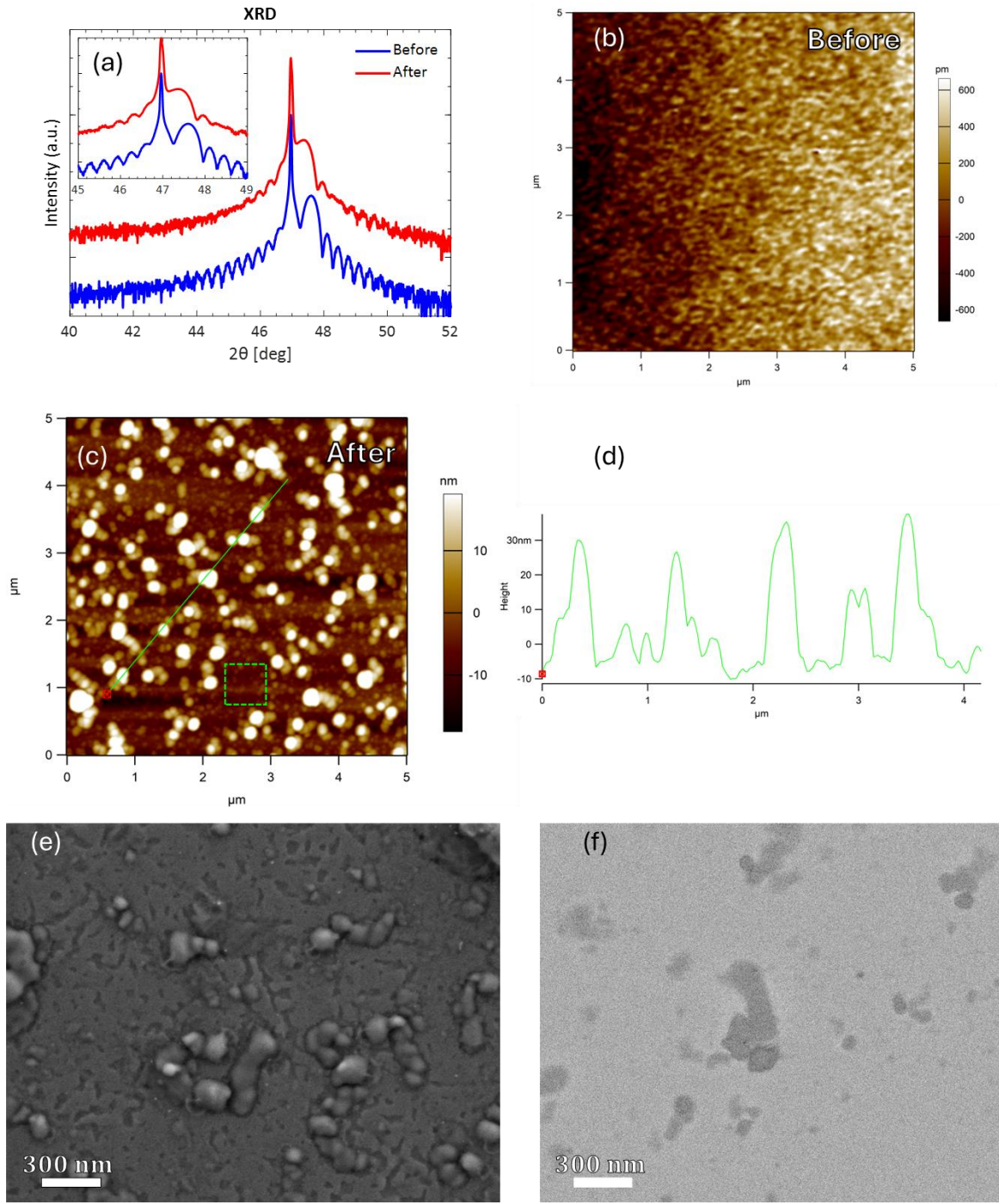

**Figure 3.** (a) XRD around the (002) Bragg peak for the sample before (blue) and after (red) the process. Inset shows a narrower scan range to emphasize differences. (b) AFM of before ($R_q$=0.16 nm), and (c) after the process. The green square indicates the region used to calculate the background roughness ($R_q$ = 2.08 nm). (d) Roughness section of the green line in (c). (e) Secondary-electron SEM (angled detector) emphasizing morphology, and (f) EsB-detector SEM highlighting Z-contrast.

To rationalize the physical mechanisms underlying the surface transformation, we next evaluate whether *chemically driven pathways* involving the reported surface-related $Sr_3V_2O_8$ and $Sr_2V_2O_7$ phases (Fig. 4a),[26–28,37] are consistent with the observed morphology and spectroscopic signatures. For each of these oxides, two competing reactions were evaluated under UHV ($\sim 1\times 10^{-10}$ Torr) and high-temperature (970 K) conditions of the *in-situ* experiment. The first set of reactions pathways involves oxygen release:

$$Sr_3V_2O_{8(s)} \rightarrow 2SrVO_{3(s)} + SrO_{(s)} + \tfrac{1}{2}O_{2(g)} \uparrow$$

$$Sr_2V_2O_{7(s)} \rightarrow 2SrVO_{3(s)} + \tfrac{1}{2}O_{2(g)} \uparrow$$

whereas the second pathway corresponds to the volatilization of vanadium pentoxide:

$$Sr_3V_2O_{8(s)} \rightarrow 3SrO_{(s)} + V_2O_{5(g)} \uparrow$$

$$Sr_2V_2O_{7(s)} \rightarrow 2SrO_{(s)} + V_2O_{5(g)} \uparrow$$

Under these conditions, we performed simplified thermodynamic estimates (see details and discussion in the Supporting Information), which do not account for the kinetics of the process. These point to the $V_2O_5$ volatilization pathway as strongly favorable over the oxygen release scenario. The $V_2O_5$ volatilization pathway ultimately yields a solid product of SrO, whose segregation is commonly reported under ambient conditions.[38] In that case, and given the AFM and HR-SEM observations, such a product would be expected to manifest as nanoscale islands (Fig. 4b). Upon subsequent air exposure, this product would be expected to undergo further reaction (Fig. 4c) to form $SrCO_3$ ($SrO_{(s)} + CO_2 \rightarrow SrCO_{3(s)}$).[38–40]

Under this segregation scenario, $SrCO_3$, having the lowest average atomic number among the candidate Sr-containing phases (Table S1), would be expected to appear darker in Z-contrast imaging, consistent with the contrast observed for the nanoscale islands in the EsB-SEM (Fig. 3f). However, the combined experimental evidence argues against a $SrO/SrCO_3$ segregation scenario as the dominant origin of the nanoscale islands. $V_2O_5$ volatilization would be expected to increase the Sr:V ratio at the surface; however, the XPS area ratio persists (Fig. S7). Moreover, *ex-situ* XPS lacks a pronounced carbonate-related feature in the C 1s region (Fig. S8), as would be expected for substantial $SrCO_3$ surface coverage. Furthermore, the absence of any additional diffraction peaks in wide-angle XRD (Fig. S9), indicates that no detectable crystalline secondary phases are formed.

Taken together, these observations indicate that, while chemically driven pathways cannot be fully excluded, they are unlikely to provide a comprehensive explanation of the observed surface morphology.

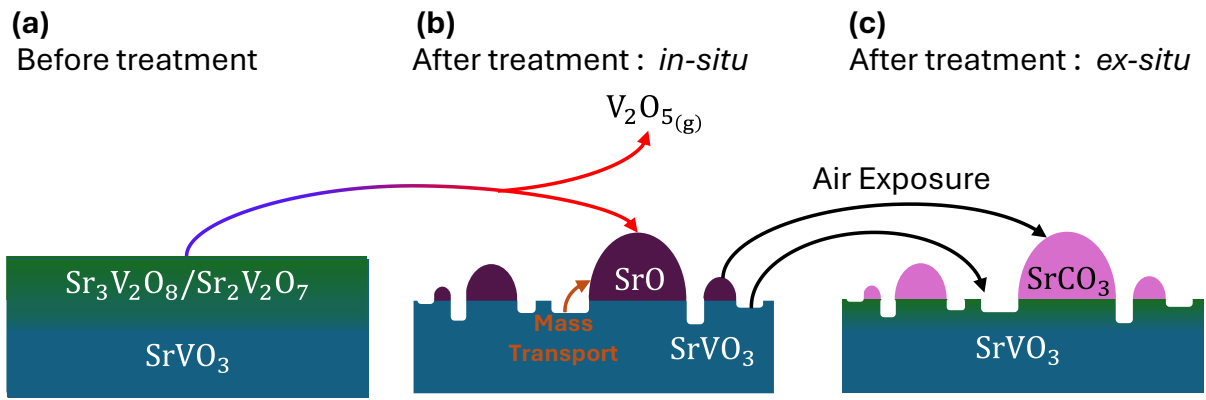

**Figure 4.** Schematic illustration of considered surface evolution pathways under *in-situ* UHV annealing and subsequent air exposure. (a) The air-exposed SVO surface is initially an oxidized, Sr-rich NSR, commonly attributed to $Sr_3V_2O_8$/$Sr_2V_2O_7$. (b) Under UHV ($\sim 10^{-10}$ Torr) and high-temperature conditions, thermodynamically favorable $V_2O_5$ volatilization leads to mass transport (marked in orange arrow) and the formation of SrO nanoscale islands and recovery of the SVO surface. (c) Upon

subsequent exposure to ambient conditions, these SrO islands can further react with $CO_2$ to form $SrCO_3$, and the exposed SVO re-oxidizes to NSR.

Given the lack of experimental support for a predominantly chemical segregation mechanism, we consider an alternative explanation based on structural surface energetics. In this scenario, changes in surface energy of the NSR could induce dewetting and island formation, thereby exposing the buried SVO layer, which could account for the predominantly $V^{+4}$ spectrum (Fig. 1b), without requiring a chemical reaction. Further study is required for pinpointing the exact chemical and morphological surface evolution.

## IV. CONCLUSIONS

We demonstrated a direct and controllable route for recovering the metallic surface of air-exposed SVO by thermally transforming the oxidized NSR under UHV. Real-time *in-situ* XPS reveals a well-defined transition from a $V^{5+}$-dominated surface to mixed valence states, with a pronounced resurgence of the $V^{4+}$ spectral signature associated with the correlated-metallic character of stoichiometric SVO. *Ex-situ* characterizations show that this chemical and electronic transformation is accompanied by mass redistribution, leading to nanoscale surface reorganization and modest lattice expansion of the SVO Thermodynamic considerations motivate evaluation of a $V_2O_5$ volatilization pathway under the experimental conditions; however, the combined experimental evidence deems this scenario unlikely and instead points toward a predominantly structural surface reorganization, with morphological dewetting remaining a viable explanation.

Together, these findings establish that predominantly $V^{4+}$ SVO surfaces can be obtained without protective capping layers, even after air exposure. The surface morphology indicates that further optimization of the process may be required for some practical purposes. This capability enables the recovery of metallic SVO surfaces after unavoidable ambient exposure, facilitating surface-sensitive spectroscopies and expanding opportunities for *ex-situ* oxide heterostructure growth, interface engineering, and oxide-electronics device integration.

## ACKNOWLEDGMENTS

We thank the Israeli Science Foundation (ISF Grant 1397/24) for funding for this work. The authors thank Dr. Maria Koifman Khristosov (Technion) for HR-SEM measurements and Dr. Pini Shekhter (TAU) for *ex-situ* XPS acquisition. We are grateful to Prof. Wayne D. Kaplan (Technion) for fruitful discussions. T. C. Back was supported by the Air Force Office of Scientific Research under Project No. FA9550-20RXCOR027. J. Ludwick was supported by the Air Force Research Labs under contract No. FA8650-16-D-5408.

## AUTHOR DECLARATIONS

### Conflict of Interest

The authors have no conflicts to disclose.

### Author Contributions

# DATA AVAILABILITY

The data that support the findings of this study are available from the corresponding authors upon reasonable request.

# Supplementary Section

# Real-time Observation of Thermal Surface Recovery in SrVO$_3$

**Amit Cohen**[1], **Jonathan Ludwick**[2,3], **Ward Yahya**[1], **Maria Baskin**[1], **Lishai Shoham**[1], **Tyson C. Back**[2], **Lior Kornblum**[*1]

[1]Andrew and Erna Viterbi Department of Electrical and Computer Engineering, Technion—Israel Institute of Technology, Haifa 32000-03, Israel

[2]Air Force Research Laboratory, WPAFB, 2179 12th Street, B652/R122, Dayton, OH 45433-7718, USA

[3]UES BlueHalo Company, 4401 Dayton-Xenia Rd, Dayton, OH 45432, USA

*Corresponding Author: liork@technion.ac.il

### Temperature Evolution and heating/cooling rate:

Temperature profile recorded by the optical pyrometer during the continuous *in-situ* XPS measurement (Fig. S1). The sample was heated from 670 K to 1052 K and then cooled back to 812 K at an average rate of ≈ 2 K/min. Minor fluctuations during the cooling stage reflect transient heater power adjustments associated with the temperature controller feedback

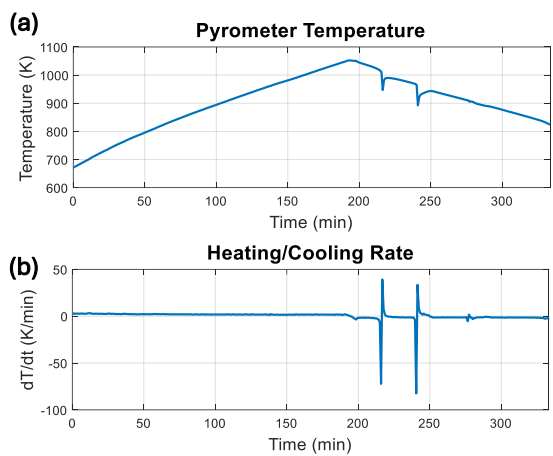

**Figure S1.** Pyrometer (a) temperature and (b) corresponding heating/cooling rate recorded during the *in-situ* XPS experiment.

### Full V 2p and O 1s core-level spectra:

The full spectral range of the V 2p and O 1s core-levels was acquired continuously during the in-situ thermal cycle (Fig. S2). Both core-levels exhibit gradual and correlated binding energy (BE) drift during heating, without any line shape change. This behavior may be attributed to changes in the electrical conductivity of the film NSR with temperature. As resistivity decreases upon heating, surface charging and contact potential effects are progressively reduced,[1] leading to an apparent shift of the measured binding energies. The common drift observed for both V 2p and O 1s confirms its non-chemical origin and reflects the evolving electrostatic conditions at the sample surface during the thermal treatment.

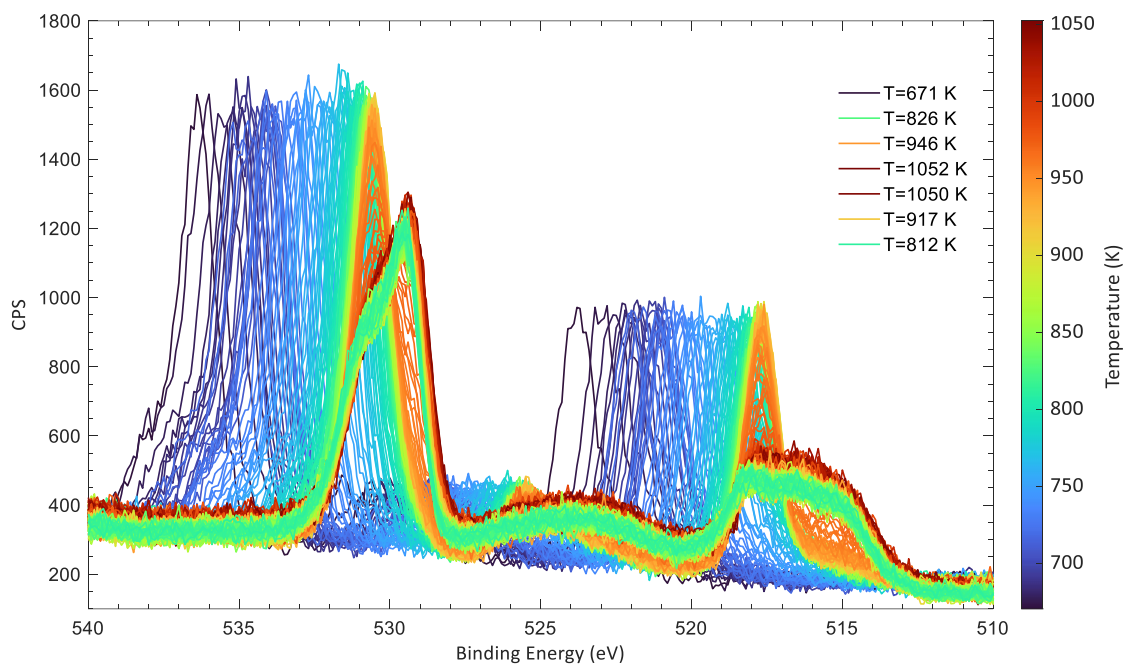

**Figure S2.** Temperature-dependent V 2p core-level spectra acquired during the *in-situ* XPS experiment.

### V 2p3/2 core-level spectra before treatment:

The V $2p_{3/2}$ core-level spectrum acquired prior to the thermal treatment (Fig. S3) is dominated by a $V^{5+}$ contribution, consistent with the presence of an oxidized NSR.[2] A weaker $V^{4+}$ component is also detected at lower BE and is attributed to the underlying stochiometric $SrVO_3$ (SVO) film, with reduced intensity due to attenuation by the NSR.

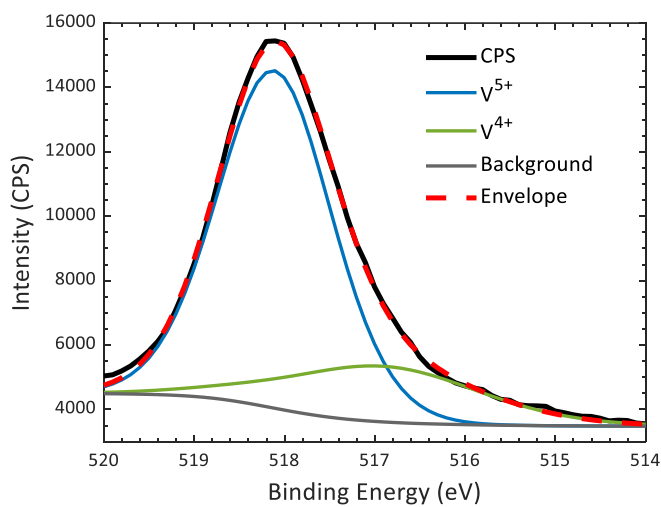

**Figure S3.** V $2p_{3/2}$ core-level spectrum acquired before the thermal treatment, shown together with the corresponding peak fit.

## V 2p core-level spectra before and after thermal treatment:

The full spectral range of the V 2p and O 1s core-levels spectra acquired before and after the *in-situ* thermal treatment (Fig. S4). After the thermal treatment, the O 1s spectrum exhibits an enhanced low BE component. This feature is attributed to the SVO film[3] and is consistent with the observed increased relative intensity of the $V^{4+}$ contribution.

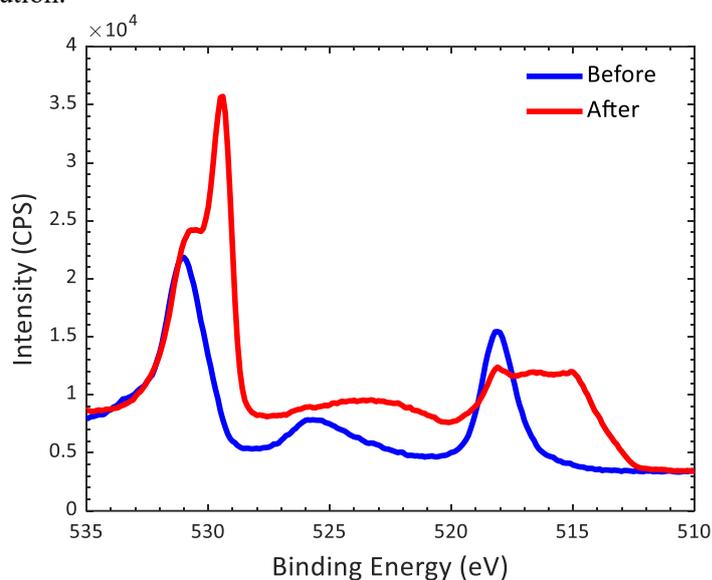

**Figure S4.** V 2p and O 1s core-levels spectra acquired before and after the in-situ thermal treatment.

## AFM particle coverage analysis:

To estimate the surface coverage of nanoscale islands formed after the thermal treatment, the AFM height map was segmented using a height threshold (Fig. S5) equal to the background roughness of the same scan (Rq = 2.08 nm, determined from the green square region shown in Fig. 3c). Using this approach, the particle coverage is approximately 27% of the scanned surface area.

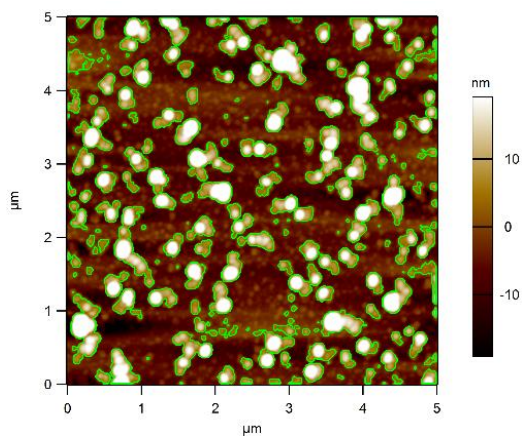

Figure S5. AFM height map of the SVO film after the in-situ thermal treatment, with the particle mask (green zones).

**SEM imaging:**

Complementary SEM imaging was performed to further characterize the surface morphology and contrast mechanisms after the thermal treatment (Fig. S6). For each imaged region, secondary-electron (SE2) and in-lens detector images were acquired from the same field of view. The SE2 detector, which collects secondary electrons at an oblique angle, primarily emphasizes surface topography, while the in-lens detector collects secondary electrons emitted close to the surface normal and provides enhanced sensitivity to compositional (Z) contrast in addition to morphology. The expected average atomic numbers ($\bar{Z}$) of the Sr–V–O phases are summarized in Table S1.

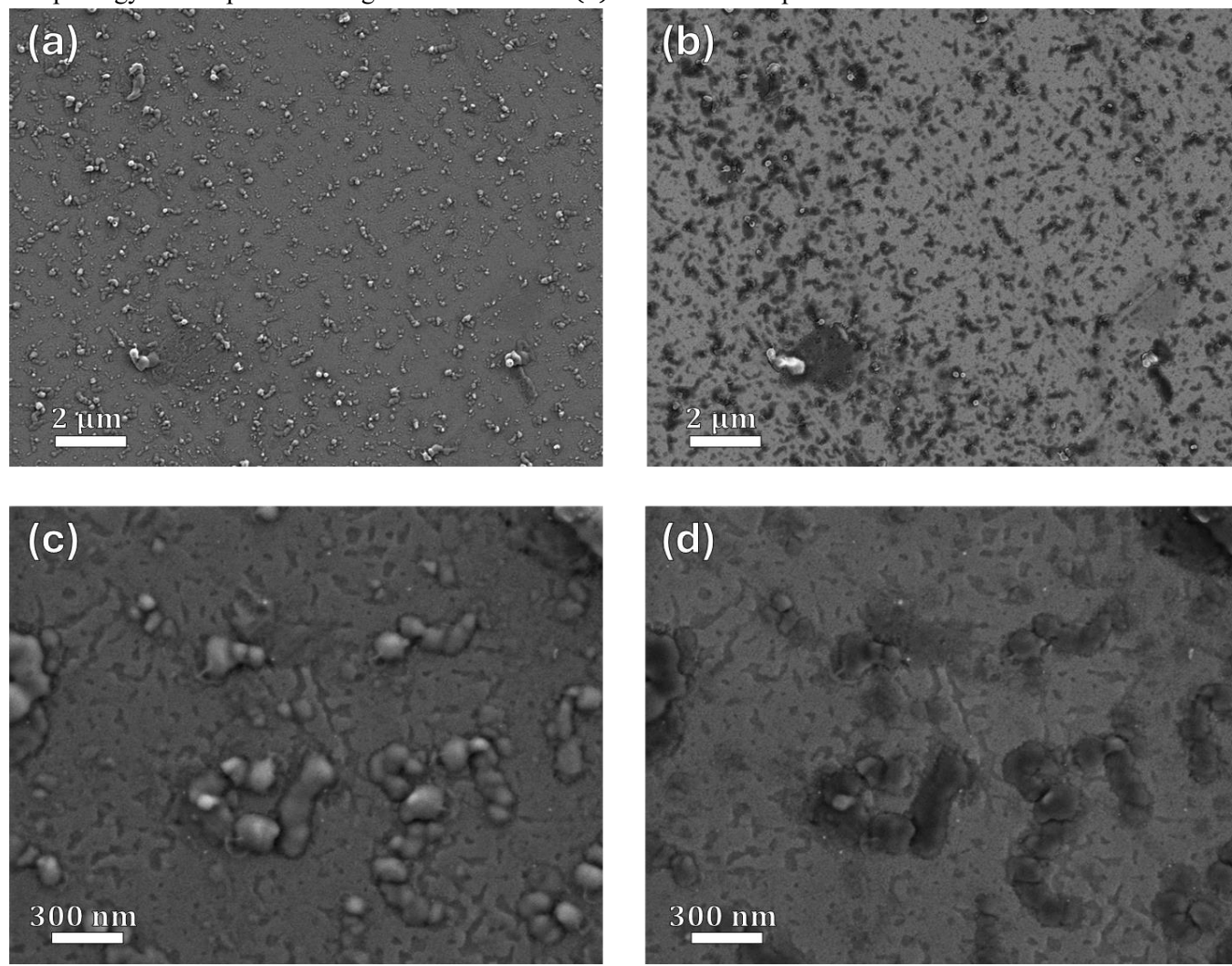

**Figure S7.** SEM images of the SVO film acquired after the thermal treatment using SE2 (left panels) and in-lens (right panels) detectors at two different magnifications: (a,b) lower magnification and (c,d) higher magnification.

Two different magnifications are shown, with a larger field of view (Fig. S6a,b) and higher magnification images (Fig. S6c,d), demonstrating that the observed islanded morphology and contrast are reproducible across length scales and representative of the overall film surface rather than a localized feature.

**Table S1.** Average atomic number ($\bar{Z}$) among Sr-containing phases considered in the surface evolution discussion.

| Phase | $SrVO_3$ | $Sr_3V_2O_8$ | $Sr_2V_2O_7$ | $SrCO_3$ |
|---|---|---|---|---|
| $\bar{Z}$ | 17 | 17.2 | 16.2 | 13.6 |

### Sr:V XPS area ratio:

To evaluate whether the thermal treatment leads to a change in surface stoichiometry consistent with $V_2O_5$ volatilization, the Sr:V XPS area ratio was extracted before and after the *in-situ* thermal process (Fig. S7). The integrated peak areas of the Sr 3d and V 2p core-levels were calculated using the same fixed energy windows (indicated by dashed lines in the figure). The resulting Sr:V ratios are 0.80 before and 0.78 after the treatment, implying that no significant enrichment of Sr occurs at the surface as a result of the thermal process.

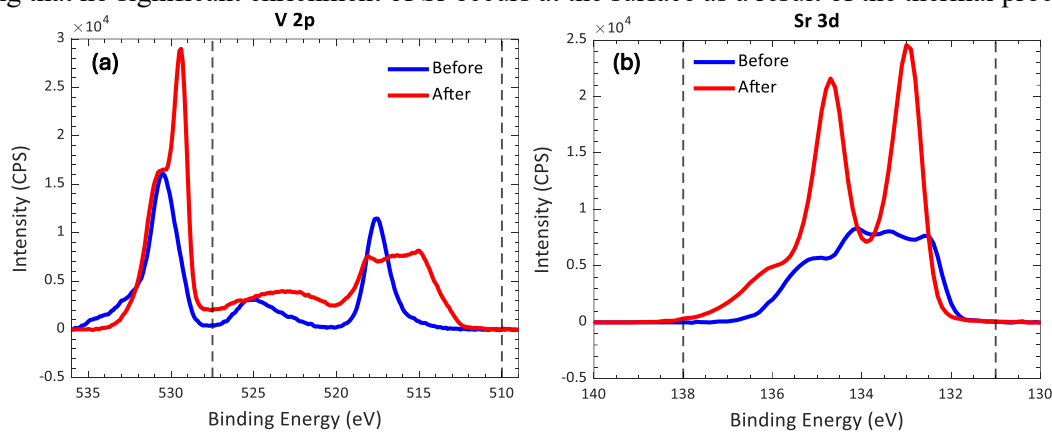

**Figure S7.** Sr:V XPS area ratio calculated from the integrated (a) V 2p and (b) Sr 3d core-level intensities before and after the in-situ thermal treatment. Integration bounds are indicated by dashed lines.

### XPS C 1s carbonate peak:

Carbonate species are expected to give rise to a characteristic C 1s feature at ~288 eV.[4] To assess whether the nanoscale islands observed after thermal treatment originate from SrO segregation followed by conversion to $SrCO_3$ upon air exposure, we compare the *ex-situ* C 1s spectrum of the post-treatment SVO film with that of a capped SVO reference sample,[3] for which Sr segregation and subsequent carbonation are expected to be suppressed (Fig. S8). The carbonate-related peak exhibits a comparable magnitude in both spectra, indicating that a substantial increase in carbonate coverage does not occur after treatment, making it unlikely that SrCO₃ constitutes a major fraction of the film surface

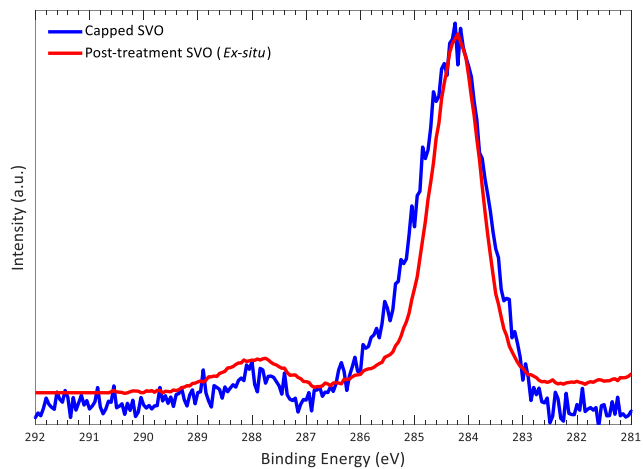

**Figure S8**. *Ex-situ* C 1s XPS spectra of post-treatment SVO compared with a capped SVO reference.[3] Carbonate-related peak is observed at 288 eV.

### Wide-angle X-ray diffraction scan after thermal treatment:

A wide-angle X-ray diffraction (XRD) scan was acquired after the in-situ thermal treatment (Fig. S9) to assess the presence of possible secondary or decomposition phases. The scan spans a broad angular range and shows only reflections associated with the LSAT substrate, including the (001), (002), and (003) peaks, together with the corresponding epitaxial SVO film reflections appearing at slightly higher angles. No additional diffraction features are observed within the measured angular range, indicating the absence of detectable secondary crystalline phases.

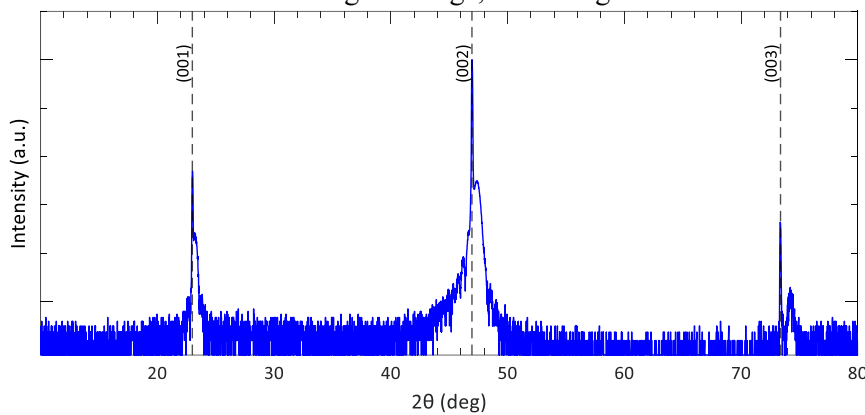

**Figure S9.** Wide-angle X-ray diffraction scan of the SrVO₃ film acquired after the in-situ thermal treatment. The vertical dashed lines mark the (001), (002), and (003) substrate reflections of LSAT.

### Thermodynamic calculations:

To assess the thermodynamic driving forces of the reaction pathways discussed in the main text, we estimated the Gibbs free energy change (ΔG) for each reaction, using

$$\Delta G = \Delta H - T\Delta S,$$

where ΔH is the enthalpy change, T is the temperature, and ΔS is the entropy change. For $V_2O_5$, $O_2$ and SrO, enthalpy and entropy values are available in the literature.[5]

For the more complex phases of $Sr_3V_2O_8$ and $Sr_2V_2O_7$, for which reliable thermodynamic data are scarce, enthalpy and entropy estimations were made. First, room temperature enthalpy and entropy of the $Sr_3V_2O_8$ and $Sr_2V_2O_7$ were approximated (Table. S1) by comparison to chemically and structurally related alkaline-earth vanadates $Ca_3V_2O_8$ and $Ca_2V_2O_7$,[6] where Ca is Sr's neighbors in the periodic table, yielding results with conservative uncertainty bounds.

Table S2. Room-temperature enthalpy and entropy estimates for Sr–V–O phases.

| Phase | $\Delta H_{RT}(MJ \cdot mol^{-1})$ | $\Delta S_{RT}(J \cdot mol^{-1}K^{-1})$ |
|---|---|---|
| $Sr_3V_2O_8$ | -3.3±0.3 | 270 ± 30 |
| $Sr_2V_2O_7$ | -2.7±0.3 | 210 ± 20 |

Next, temperature-dependent corrections from room temperature (300 K) to the experimental temperature (970 K) were estimated using heat-capacity ($C_p$) integration at the pressure of experiment (~1×10⁻¹⁰ Torr):

$$\Delta H = \int_{300}^{970} C_p dT; \ \Delta S = \int_{300}^{970} \frac{C_p}{T} dT$$

Heat capacities were estimated using the Neumann–Kopp law, which is applicable at temperatures well above the Debye temperature. The pressure dependence of $\Delta G$ enters only through the gaseous species in the reactions. The enthalpy of sublimation of vanadium pentoxide was estimated based on literature data.[7,8]

Using these inputs, the Gibbs free energy changes of the proposed reactions under the experimental *in-situ* conditions (970 K, ~1×10⁻¹⁰ Torr) were evaluated (Table. S2).

Table S3. Reaction pathways and calculated Gibbs free energy changes at 970 K and UHV pressure.

| Reaction | $\Delta G \ (kJ \cdot mol^{-1})$ |
|---|---|
| $Sr_3V_2O_{8(s)} \rightarrow 2SrVO_{3(s)} + SrO_{(s)} + \tfrac{1}{2}O_{2(g)} \uparrow$ | 10 ± 30 |
| $Sr_3V_2O_{8(s)} \rightarrow 3SrO_{(s)} + V_2O_{5(g)} \uparrow$ | −280 ± 30 |
| $Sr_2V_2O_{7(s)} \rightarrow 2SrVO_{3(s)} + \tfrac{1}{2}O_{2(g)} \uparrow$ | 10 ± 30 |
| $Sr_2V_2O_{7(s)} \rightarrow 2SrO_{(s)} + V_2O_{5(g)} \uparrow$ | −280 ± 30 |

These results show that reaction pathways involving the formation and volatilization of $V_2O_5$ are strongly favored over oxygen-release mechanisms under the experimental conditions. While the oxygen-release reactions are marginal within uncertainty, the $V_2O_5$-forming reactions are stabilized by several hundred $kJ \cdot mol^{-1}$, identifying $V_2O_5$ volatilization as the thermodynamically dominant pathway.